\documentstyle[prl,aps,twocolumn,eqsecnum,epsfig]{revtex}
\global\firstfigfalse 
\begin{document}
\draft
\preprint{AZPH-TH/97-07}
\twocolumn[\hsize\textwidth\columnwidth\hsize\csname @twocolumnfalse\endcsname
\title{Partonic Picture of Nuclear Shadowing at Small $x$}
\author{Zheng Huang, Hung Jung Lu and Ina Sarcevic}
\address{Department of Physics, University of Arizona, Tucson, AZ 85721}
 
\date{May 6, 1997}
\maketitle
 
\begin{abstract}
We investigate the nuclear shadowing mechanism in the context of
perturbative QCD and the Glauber multiple scattering model. 
Using recent HERA data on nucleon structure function at small $x$,
we put stringent constrains
on the nucleon gluon density in the double-logarithm approximation.
We suggest that the scaling violation 
of the nucleon structure function in the region of
small $x$ and semihard scale $Q^2$ can be reliably
described by perturbative QCD which
 is a central key to the
understanding of the scale dependence of the nuclear shadowing effect.
Our results indicate that while the shadowing of
the quark density arises from an interplay between the ``soft''
and semihard QCD processes, the gluon shadowing is 
largely driven by a perturbative shadowing mechanism. We demonstrate that
the gluon shadowing is a robust phenomenon at large $Q^2$ and can
be unambiguously predicted by perturbative QCD.
\end{abstract}

]
 
\section{Introduction}
The theoretical studies of semihard processes play a increasingly
 important role in ultrarelativistic
heavy-ion collisions at $\sqrt{s}\geq 200$ GeV in describing the global
collision features such as the particle multiplicities 
and transverse energy distributions. 
These semihard processes can be reliably calculated in the framework of
perturbative QCD and they are crucial for determining the
initial condition for the possible formation of
a quark-gluon plasma.  Assuming the validity of the
factorization theorem in perturbation theory,
it is essential
to know the parton distributions in nuclei in order to 
compute these processes.  For example, to
calculate the minijet production for $p_T\sim 2$ GeV or the charm
quark production 
in the central rapidity region ($y=0$), one needs to know the nuclear
gluon structure function, $xg_A(x,Q^2)$, at $Q\sim p_T$ or $m_c$ and
$x=x_{\rm Bj}=2p_T/\sqrt{s}$ or $2m_c/\sqrt{s}$. For $\sqrt{s}\geq 200$
GeV, the corresponding value of $x$ is $x\leq 10^{-2}$. Similarly, the
knowledge of nuclear quark density $xf_A(x,Q^2)$ is needed for
the Drell-Yan lepton-pair production process.

The attenuation of quark density  has been firmly established
experimentally in deeply-inelastic lepton scatterings (DIS) from nuclei 
at CERN \cite{cern} and Fermilab \cite{fermilab} 
in the region of small values of the Bjorken variable $x=Q^2/2m\nu$ , where
$Q^2$ is the four-momentum transfer squared, $\nu$ is the energy loss
of the lepton in the lab frame and $m$ is the nucleon mass.
 The data, taken over a
wide kinematic range  $10^{-5}<x<0.1$ and 0.05 ${\rm GeV}^2 <
Q^2< 100$ GeV$^2$, show a systematic reduction of nuclear structure function
$F_2^A(x,Q^2)/A$ with respect to the free nucleon structure function
$F_2^N(x,Q^2)$. There are also some indications of nuclear gluon shadowing 
from the analysis of J/$\Psi$ suppression in hadron-nucleus
experiments \cite{glushdw}. Unfortunately, the extraction of nuclear
gluon density is not unambiguous since it involves the evaluation of
initial parton energy loss 
and final state interactions \cite{fs}.
These measurements present a tremendous challenge to the theoretical
models of nuclear shadowing (for recent reviews, see \cite{rev}). 

The partonic description of nuclear shadowing phenomenon has been
extensively investigated  in literatures in the context of a
recombination model \cite{glr,mq,eqw} and a Glauber multiple scattering
theory \cite{frankfurt,lu,zakharov,mueller,agl}. These approaches predict
a modified QCD evolution equation for the nuclear structure function
in contrast to a Gribov-Lipatov-Altarelli-Parisi (GLAP) evolution equation 
\cite{glap} for nucleon. Qualitatively, the scale ($Q^2$) evolution of the
nuclear structure function in the modified equation
is slower than the nucleon case, leading to a perturbative mechanism
for the depletion of the parton density at large momentum 
transfer. The quantitative prediction of nuclear shadowing 
at large $Q^2$ in general depends on the initial value of shadowing
at small momentum transfer where perturbative calculation breaks
down and some non-perturbative model or experimental input is needed.
The so-called vector-meson-dominance (VMD) model \cite{vmd} has been quite
successful in predicting the quark shadowing at low $Q^2$ where
the experimental data are rich \cite{cern,fermilab}.
Presently, there  is no viable non-perturbative
model for the gluon case and there is very little experimental
information on the initial gluon shadowing value at some low $Q_0^2$.
This, in practice, would make the prediction on gluon shadowing 
very uncertain.

In this paper, however, we suggest that there exists a typical scale 
$Q_{\rm SH}^2\simeq 3$ GeV$^2$ in 
the perturbative evolution of the nuclear
gluon density  beyond which the nuclear gluon shadowing can be 
unambiguously predicted in the context of perturbative QCD. The
so-called ``semihard'' scale $Q_{\rm SH}^2$ is determined by the
strength of the scaling violation of the nucleon gluon density in the
small-$x$ region at which the anomalous dimension, 
$\gamma \equiv \partial \ln xg_N/\ln Q^2$, is of $O(1)$. We shall
demonstrate that as $Q^2$ approaches $Q_{\rm SH}^2$,
the gluon shadowing ratio rapidly approaches a unique
perturbative value determined  by the Glauber multiple scattering theory. 
For $Q^2>Q_{\rm SH}^2$,  the  nuclear
gluon density, in contrast to the quark case,
  is almost independent of the initial 
distribution at $Q_0^2$ and its evolution is  mainly governed by a 
GLAP equation. The suggested phenomenon is a consequence of the
singular growth of the gluon density in the small-$x$ region and 
the perturbative shadowing caused by the coherent 
parton multiple scattering.

In particular, 
we shall present a quantitative study of the scale dependence of
the nuclear shadowing phenomenon for both the quark and the gluon 
distributions based on the Glauber diffractive approximation. 
A similar study has been performed by Eskola in the 
recombination model \cite{eskola1}. Our focus is the 
sensitivity of the nuclear
shadowing at large $Q^2$ to the initial conditions.
To determine the gluon content of the nucleon, we shall use a leading
double logarithm approximation (DLA) where 
the scaling violation of the quark structure function
$F_2^N$ is entirely caused by a $g\rightarrow q \bar q$ 
splitting. The 
agreement between our results based on the DLA and the measured 
values of $F_2^N$ indicates that the perturbative evolution sets
in as early as $Q^2\geq 0.8$ GeV$^2$ and the scaling violation of gluon
structure function in the semihard region 
 can be reliably described by the DLA. 
This provides a basis for studying both the nuclear quark 
and gluon densities in the Glauber approach 
where the separation of $q\bar q$ or $gg$ pair in the DLA is 
considered frozen during its passage in the nucleus.
The reason for not using the available parton distributions
for a nucleon such as those of 
MRS, CTEQ or GRV parameterization \cite{mrs}
 is that the nuclear structure function 
obtained in our Glauber approach does not contain a full GLAP kernel, as
it will become clear below. Since the shadowing ratio depends on both
the nuclear and the nucleon structure functions, they should be 
consistently calculated
in the same approximation. We expect that 
the calculated ratio, however, has a
validity beyond the DLA.

We shall demonstrate that the nuclear shadowing for the quark distribution
is a leading twist effect due to the interplay between the ``soft''
and the ``semihard'' QCD processes, while for the
gluon distribution, the shadowing is largely driven by the
perturbative shadowing mechanism at small $x$. 
Beyond the semihard scale, both quark and
gluon shadowing ratios are shown
to evolve slowly.  
We discuss the $x$-dependence
of the shadowing ratio and the non-saturating feature of the perturbative
shadowing. As a by-product of the Glauber approach, we also study the 
nuclear geometry dependence of the shadowing ratio, which
cannot be addressed in the recombination model such as the GLR
equation.  Since the shadowing effect
is intimately related to the effective number of scatterings, it should
vary strongly as one changes the impact parameter. This leads to the
breakdown of the factorization of the nuclear parton density
 into the product of 
an impact parameter averaged parton distribution and the nuclear thickness
function.

The paper is organized as follows. In section \ref{sec:pic}
we discuss the physical picture of the nuclear
shadowing phenomenon. It can be regarded as a heuristic introduction
to the central idea of our approach.  
In section \ref{sec:f2n}, we calculate the nucleon structure function
$F_2^N(x,Q^2)$ for large range of $x$ and $Q^2$
using a gluon density obtained in the DLA.  We determine the
initial condition for $F_2^N$ and for $xg_N$ by 
fitting the $Q^2$-dependence of
$F_2^N$ at small value of $x$.
The gluon density in a 
nucleon is crucial to the nuclear shadowing effect.
In section \ref{sec:glb}
we give a brief introduction on the Glauber multiple scattering 
model and discuss the dependence of the nucleon density on the
coherence length of the virtual photon.
The nuclear shadowing effect for $F_2^A$ is calculated in
section \ref{sec:fa} and the initial shadowing is parameterized in order
to fit the available experimental data at the measured $Q^2$ values.
In section \ref{sec:ga}, the gluon shadowing is computed using the Glauber
formula derived by Mueller \cite{mueller}
and Ayala F, Gay Ducati and Levin \cite{agl}. Section \ref{sec:impact} is
devoted to the discussion of the impact parameter dependence 
 of the nuclear shadowing. We conclude our findings 
in section \ref{sec:conc}.

\section{Physical Picture of  Nuclear Shadowing}
\label{sec:pic}
At low $Q^2$ ($< 1$ GeV$^2$) in the
deeply-inelastic scattering, the interaction of the virtual photon
with the nucleons in the rest frame of the target is most naturally 
described by a vector-meson-dominance (VMD) model \cite{vmd}, in which 
the virtual photon interacts with the nucleons via its hadronic fluctuations,
namely the $\rho$, $\omega$ and $\phi$ mesons. The shadowing of 
the (sea) quark density is caused by the coherent multiple scatterings
of the vector meson off the nucleons with a large hadronic cross section
$\sigma_{VN}$, which can be most conveniently studied in the 
Glauber diffractive approximation model \cite{glauber}.
The $x$-dependence of the shadowing ratio, 
$R(x,Q^2)=F_2^A(x,Q^2)/AF_2^N(x,Q^2)$, is explained by the coherent length
of the virtual photon $l_c\simeq 1/2mx$. For $x<10^{-2}$, $l_c>R_A=4\sim 5$ 
fm for large nuclei, the vector meson interacts coherently with all nucleons
inside the nucleus resulting in a uniform (in $x$)
 reduction of the total cross section,
while for $x>10^{-2}$, $l_c<R_A$ the vector meson can only interact
coherently with a fraction of all the nucleons and shadowing is 
reduced.  The VMD model in general predicts much less scaling
violation of $F_2^N(x,Q^2)$ than it has been
 observed in the small $x$ region
and a rapid disappearance of the nuclear
shadowing for $Q^2\gg m_V^2$. Although a generalized VMD model which includes
higher resonances may improve the situation somewhat, it is
generally believed that the nuclear shadowing in the VMD model is a higher
twist effect, which is in contraction with the experimental evidence that
the $R(x,Q^2)$ has a rather weak $Q^2$ dependence \cite{cern,fermilab}. 

At high $Q^2$ ($>1\sim 2$ GeV$^2$), the virtual photon can penetrate the
nucleon and probe the partonic degrees of freedom where
 a partonic interpretation based on perturbative QCD is most relevant
in the infinite momentum frame.
It is well known that the $Q^2$-dependence, the so-called scaling violation
of the nucleon structure function $F_2^N(x,Q^2)$ can be well accounted
for by the Gribov-Lipatov-Altarelli-Parisi (GLAP) evolution equation 
\cite{glap} given some non-perturbative initial condition $F_2^N(x,Q_0^2)$
\cite{mrs}. 
In the small $x$ region, the GLAP equation predicts a strong scaling
violation of $F_2^N(x,Q^2)$ due to the strong growth of the gluon
density in the nucleon. In fact, a dominant perturbative evolution
implies that the growth of $F_2^N(x,Q^2)$ in $\ln Q^2$ at fixed, small $x$
is closely related to the growth in $\ln 1/x$ at fixed $Q^2$, i.e.\ the
so-called double scaling violation
observed in the DIS experiments for nucleon, 
most transparent in the Leading-Double-Logarithm
Approximation (DLA) of the gluon structure function. 
At extremely small $x$, the resummation of large $\ln 1/x$ is necessary
leading to the BFKL evolution equation \cite{bfkl} which may also be 
derived in a Weiszacker-Williams classical field approximation \cite{mv}. 
It is not clear
when the BFKL hard pomeron will become relevant since the GLAP evolution
seems able to describe most of kinematical region that current experiments
have probed so far. As we shall show,
the main feature of the $Q^2$-dependence of the nuclear shadowing is 
intimately related to this strong scaling violation in the small-$x$ region.

If the nuclear structure function follows the same perturbative 
GLAP evolution, given
an initially shadowed $F_2^A(x,Q_0^2)$,
the shadowing ratio $R(x,Q^2)$ as $Q^2$ increases will quickly
approach $1$ because of the strong scaling violation 
$F_2^A(x,Q^2)\gg F_2^A(x,Q_0^2)$. This would imply that the shadowing
is purely non-perturbative  and thus a higher twist effect. 
The key observation is that in the infinite momentum frame, because of
the high nucleon and parton densities, the quarks and gluons that belong
to different nucleons in the nucleus will recombine and annihilate, leading to
the so-called recombination effect first suggested by Gribov, Levin and
Ryskin \cite{glr} and later proven by Mueller and Qiu \cite{mq}. 
In the target rest frame, the virtual photon interacts with nucleons
via its quark-antiquark pair ($q\bar q$) color-singlet 
fluctuation \cite{lu}. If the coherence
length of the virtual photon is larger than the radius of the nucleus,
$l_c>R_A$, the $q\bar q$ configuration interacts
coherently with all nucleon with a cross section given by the
color transparency mechanism for a point like color-singlet configuration
\cite{plc}. That is, the cross section is proportional to the transverse
separation squared, $r_t^2$, of the $q$ and $\bar q$. In the double logarithm
approximation (DLA), the cross section can be expressed through
the gluon structure function for a nucleon
\begin{eqnarray}
\sigma_{q\bar q N} \propto r_t^2\alpha_sx'g_{\rm DLA}(x',1/r_t^2)\; , \label{1}
\end{eqnarray}  
where $x'=M^2/2m\nu$ and $M$ is the invariant mass of $q\bar q$ pair.
The total cross section $\sigma_{q\bar q A}$ can be calculated in the
Glauber eikonal approximation, 
\begin{eqnarray}
\sigma_{q\bar q A}=\int d^2b 2\{ 1-e^{-\sigma_{q\bar q N}T_A(b)/2}\}\; ,
\label{2}
\end{eqnarray}
where $b$ is the impact parameter and $T_A(b)$ is the nuclear
thickness function. The nuclear cross section is therefore reduced  
when compared to the simple addition of free nucleon cross sections.

However, it is not clear that (\ref{1}) and (\ref{2})
 will provide a 
satisfactory explanation of the non-vanishing shadowing effect at 
large $Q^2$. Although for a given $Q^2$, all $q\bar q$ configurations
with the separation size $r_t^2>1/Q^2$ are possible, 
the virtual photon wavefunction
$|\psi_{\gamma ^*}|^2\sim 1/r_t^2$ favors the smallest value of 
$r_t^2\simeq 1/Q^2$
where the cross section is small and no significant shadowing would
be expected. The resolution to this problem comes from the fact that
the nucleon structure function exhibits a  strong
scaling violation in the small $x$ region
at the moderate (semihard) momentum transfer
$0.8\;\; {\rm GeV}^2<Q^2<Q^2_{\rm SH}$ where $Q^2_{\rm SH}\simeq 3$ GeV$^2$.
As we shall show, the strong scaling violation in this region
can be understood in the context of perturbative QCD. 
The gluon density in (\ref{1}) grows rapidly with $1/r_t^2$ in the
small $x$ region, which can compensate the inherent ``smallness''
of the cross sections of ``hard'' QCD processes. As a consequence,
the semihard processes can compete successfully with the ``soft'' processes
as described in the VMD model and continue to provide the shadowing mechanism 
up to some moderately high value of $Q^2\sim Q^2_{\rm SH}$. 
In terms of the anomalous dimension $\gamma (Q^2)=d\ln xg_N/d\ln Q^2$,
the cross section in (\ref{1}) may be effective expressed as
\begin{eqnarray}
\sigma_{q\bar q N} \propto \frac{(Q^2)^{\gamma}}{Q^2}\; .
\end{eqnarray}
The large anomalous dimension, $\gamma >1$, even at moderately
large $Q^2$ delays the rapid falling of the cross section due to
the ``smallness'' of the configuration size,
 as first argued by Gribov, Levin and Ryskin in \cite{glr}.
We refer to this as the perturbative shadowing mechanism as it can be computed
in perturbative QCD.
At even higher
values of $Q^2$, $Q^2>Q^2_{\rm SH}$, 
the scaling violation becomes weak (logarithmic),
and the perturbative shadowing mechanism ceases to be effective. In this case,
both the nucleon and the nuclear structure function evolve slowly at
the same rate.
The shadowing ratio developed by both perturbative and non-perturbative
mechanisms in the relatively low-$Q^2$ region persists in the high-$Q^2$ region
but tends to $1$ rather slowly as $Q^2\rightarrow \infty$.

In contrast to the VMD model, the $x$-dependent cross section 
$\sigma_{q\bar q N}$ predicts an additional $x$-dependence of the
shadowing ratio apart from the  coherence length dependence in the
perturbative region. As a result, one does not in general expect 
the saturation of the shadowing ratio as $x$ decreases below the coherence
limit $x<10^{-2}$. On the other hand, as $Q^2$ decreases, the gluon
distribution changes from a singular small-$x$ behavior (hard
pomeron regime) to the less singular one (soft pomeron regime). The 
shadowing ratio  becomes flatter as $Q^2$ decreases. The
observed saturation of the shadowing ratio in the small-$x$ region 
\cite{fermilab} is thus a non-perturbative, low-$Q^2$ feature.

Similarly, the nuclear gluon density may be studied 
by considering a deeply-inelastic scattering of a
virtual colorless probe (e.g. the virtual ``Higgs boson'')
 with the nucleus where the interaction proceeds
via a gluon pair ($gg$) component of the virtual probe. This
has been introduced by Mueller \cite{mueller}
and recently studied by Ayala F, Gay Ducati and Levin (AGL) \cite{agl}.
In contrast to the quark distribution, the gluon structure
functions for nucleon and nucleus are poorly known, especially
in the small-$x$ region. They are both of extreme importance for the
semihard QCD processes at high energies. Most of our knowledge 
about the
gluon distribution comes from the theoretical expectations based
on perturbative QCD constrained by the scaling violation of $F_2^N$
and the prompt photon experiments \cite{mrs}. 
The nuclear gluon density may also be studied in the Glauber
multiple scattering model where $\sigma_{q\bar q N}$ in (\ref{2})
is to be replaced by $\sigma_{gg N}$. Simple color group representation
argument relates these two cross sections by 
$\sigma_{gg N}=9\sigma_{q\bar q N}/4$. However, what makes the gluon
case very different from the quark case is the unknown initial 
(non-perturbative) gluon density in the nucleus. There is very little 
experimental information on the gluon shadowing at low $Q^2$ and small $x$.
The VMD model is not applicable here since the low-lying meson spectroscopy
for $gg$-like states is poorly known. This would make this approach
very uncertain in predicting the amount of gluon shadowing in the nucleus.

However, as we shall show in this paper, the larger cross section
$\sigma_{gg N}$ implies an even 
stronger scaling violation of gluon density than the quark case, 
making the perturbative shadowing a dominant
mechanism. As a consequence, the shadowing ratio is largely determined
by the perturbative calculation at $Q^2=Q^2_{\rm SH}$. That is, the large
anomalous dimension of gluon distribution function for a nucleon drives
the shadowing ratio rapidly to approach the perturbative value at moderately
large $Q^2=Q^2_{\rm SH}$ and the influence of an 
initial shadowing at low $Q^2$ becomes relatively unimportant. 
For $Q^2>Q^2_{\rm SH}$, the scaling violation
becomes weaker, the shadowing ratio determined by
the perturbative shadowing mechanism at the semihard scale persists and
evolves slowly. The $x$-dependence of the shadowing ratio can be
also predicted to be non-saturating, a generic feature of the 
perturbative shadowing.
We therefore suggest that the nuclear gluon shadowing
at $Q^2>Q^2_{\rm SH}\simeq 3$ GeV$^2$ and at small $x$ 
can be predicted in the context of
perturbative QCD. The prediction is based on the scaling violation
of the gluon structure function and the existence of the
perturbative shadowing in the Glauber multiple scattering model.

\section{Lab Frame Description of Nucleon Structure Function}
\label{sec:f2n}
Since the nuclear shadowing ratio depends sensitively on the nucleon
structure function, we shall develop a formalism to calculate both
the nuclear and the nucleon structure functions simultaneously
in order to reduce the uncertainty due to the nucleon structure function.
This prevents us from using the well-known
parametrizations such as MRS, CTEQ and GRV \cite{mrs} which are
based on the full next-to-leading-order GLAP equation 
because the nuclear structure function is
 not derived in the same formalism. We stress that the shadowing is
 only a few tens percentage effect, thus the uncertainty in the nucleon
structure function at the same level can cause a large effect on the shadowing
ratio.

\subsection{Quark Distribution}
\begin{figure}
\centerline{\epsfig{figure=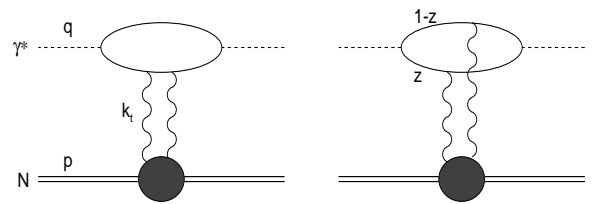,width=3in,height=1in}}
\vskip 0.1in
\caption{The perturbative diagrams giving rise to the
$q\bar q$ pair fluctuation.}
\label{fig:1}
\end{figure}

Consider a description of deeply-inelastic lepton scattering off a nucleon
in the laboratory frame. The inclusive cross section can be 
expressed via the virtual photon-nucleon total cross section. The cross
section for the absorption of a virtual photon in the small $x$ region
is dominated by the scattering off the sea-quark $q\bar q$ pair. The
generic perturbative QCD diagrams giving rise to the $q\bar q$ fluctuation
are shown in Fig.\ \ref{fig:1}. The CM energy of the incoming virtual
photon at small $x$ is 
\begin{eqnarray}
s=(q+p)^2\simeq 2p\cdot q=\frac{Q^2}{x}\; ,
\end{eqnarray}
where $q$ and $p$ are the four-momenta of the photon and the target nucleon.
Thus a region of small $x$ corresponds to a high energy scattering process 
at fixed $Q^2$.

The imaginary part of the amplitudes in Fig.\ \ref{fig:1} is related
to the photoabsorption cross section which has been calculated by
Nikolaev and Zakharov \cite{zakharov} assuming that the size of the  
$q\bar q$ pair is frozen in the scattering process and that the one-gluon 
exchange process dominates. This is nothing but the so-called double logarithm
approximation (DLA) \cite{glr,dla}. 
The transverse cross section can be cast into an impact 
parameter representation 
\begin{eqnarray}
\sigma (\gamma^*N)=\int_0^1dz \int d^2{\bf r}|\psi (z,{\bf r})|^2
\sigma_{q\bar qN}({\bf r})\; , \label{eq:sig}
\end{eqnarray}
where $z$ is the Sudakov variable, defined to be the fraction of 
the $q\bar q$ pair momentum carried by one of the (anti)quarks,
${\bf r}$ is the variable conjugate to the transverse momentum
of (anti)quark of the pair, representing the transverse size of the
pair. $\psi (z,{\bf r})$ can be interpreted as the wave function
of the $q\bar q$ fluctuations of virtual photon. Thus, 
$|\psi (z,{\bf r})|^2$ serves as the probability of finding 
a $q\bar q$ pair with a separation ${\bf r}$ and a fractional
momentum $z$. In an explicit form
\begin{eqnarray}
|\psi (z,{\bf r})|^2=\frac{6\alpha_{\rm em}}{(2\pi )^2}
\sum_1^{N_f}e_i^2P(z)a^2K_1(ar)^2\; ,\label{eq:wf}
\end{eqnarray}
where $a^2=Q^2z(1-z)$ and $K_1$ is the modified Bessel function.
The factor $P(z)=z^2+(1-z)^2$ is the universal splitting function 
of gauge boson into the fermion pair. 
Since $K_1$ is a rapidly falling function, the integral in (\ref{eq:sig})
has important contributions only in the region $ar\ll 1$.
In this case, one has
\begin{eqnarray}
|\psi (z,{\bf r})|^2\rightarrow \frac{6\alpha_{\rm em}}{(2\pi )^2}
\sum_1^{N_f}e_i^2\frac{1}{r^2}\; ,\label{eq:r}
\end{eqnarray}
where the expansion $K_1(ar)^2\rightarrow 1/a^2r^2$ has been used.
The photon wave function favors a $q\bar q$ pair with small 
transverse separation.

The cross section for the high energy interaction of a small-size
$q\bar q$ configuration with the nucleon can be unambiguously 
calculated in QCD in the region of small $x$ by applying the QCD factorization
theorem. In the DLA, the result is \cite{strikman}
\begin{eqnarray}
\sigma_{q\bar qN}({\bf r})=
\frac{\pi^2}{3}r^2\alpha_s(Q'^2)x'g_N^{\rm DLA}(x',Q'^2)
\; , \label{eq:ct}
\end{eqnarray}
where $xg_N^{\rm DLA}$ is the gluon distribution of the nucleon calculated
in the DLA. The fact that the cross section is proportional to the
size of the point-like configuration is the consequence of the color
transparency in QCD. Although the $r^2$ factor in (\ref{eq:ct})
decreases fast as one goes to the short distance, 
the singular behavior of the
wave function in (\ref{eq:r}) and the strong
scaling violation of  $xg_N^{\rm DLA}$ in the small-$x$ region as 
$r$ decreases can compensate the smallness of cross section due to
the color transparency. As a result, the soft and hard physics compete
in a wide range of $Q^2$.

The $x'$ and $Q'^2$ in (\ref{eq:ct}) are related to $z$, $x$ and $r^2$
 as follows. The
invariant mass squared of the $q\bar q$ pair is 
\begin{eqnarray}
M^2\simeq \frac{k_t^2}{z(1-z)}\; ,
\end{eqnarray}
where $k_t^2\simeq 1/r^2$ is the transverse momentum squared
 of the (anti)quark. The virtuality of the gluon is then
$Q'^2=(k_t-(-k_t))^2=4/r^2$ while $x'$ is given by
\begin{eqnarray}
x'=\frac{M^2}{2m\nu}=\frac{k_t^2}{z(1-z)2m\nu}\simeq 
\frac{x}{a^2r^2}\; ,\label{eq:xp}
\end{eqnarray}
where $\nu$ is the energy loss of the lepton. The condition 
$ar<1$ implies $x'>x$. Since $r^2>4/Q^2$, the integration region in
$z$ is limited by $a^2r^2=Q^2r^2z(1-z)<1$. 
Because the integrand in (\ref{eq:sig}) is symmetric
under $z\Leftrightarrow 1-z$, one has 
$\int_0^1dz\rightarrow 2\int_0^\epsilon dz$ where $\epsilon$ is the solution
to $ar=1$ or $x'=x$. 

At small $x$, the free nucleon structure function can be written 
in terms of the
virtual photon-nucleon cross section
\begin{eqnarray}
F_2^N(x,Q^2)=\sum_ie_i^2xf_N(x,Q^2)=
\frac{Q^2\sigma (\gamma^*N)}{4\pi^2\alpha_{\rm em}}
\; ,\label{eq:vir}
\end{eqnarray}
where $xf_N(x,Q^2)$ is the (sea) quark distribution function. 
Substituting (\ref{eq:sig}) into the above equation
and changing the integration variable $z$ into $x'$
 in (\ref{eq:sig}) $dz=-xdx'/(x'^2Q^2r^2)$, one obtains
\begin{eqnarray}
f_N(x,Q^2) & = & f_N(x,Q_0^2)+\frac{3}{4\pi^3}\int_x^1\frac{dx'}{x'^2}
\nonumber \\
& & \times
\int_{4/Q^2}^{4/Q_0^2}\frac{dr^2}{r^4}\sigma_{q\bar qN} 
(x',r^2)\; , \label{eq:fn}
\end{eqnarray}
where $\sigma_{q\bar qN}$ is given in (\ref{eq:ct}). 
We have split the integration on $r^2$ into the region 
$4/Q^2<r^2<4/Q_0^2$ where the perturbative calculation is 
valid and the region $r^2>4/Q_0^2$ where the perturbation theory 
breaks down and the cross section is not calculable. The non-perturbative
part is taken as an initial condition $f_N(x,Q_0^2)$ which 
should be determined by the experimental measurement. 
However, it is worthwhile to emphasize that because of the
exponential cutoff of the large $r^2$ by $K_1(ar)^2$ in 
(\ref{eq:sig}), the integration in $r^2$ is infrared safe.
Even if one substitutes the small-$r$ cross section into 
(\ref{eq:fn}), the integral is only logarithmically divergent
$\sim \ln Q_0^2$, in contrast to the usual factorization scheme
of the hard process.
This allows a smooth extrapolation of the ``semihard'' physics into
the ``soft'' physics. Thus one can choose a rather small 
cutoff $Q_0^2$ in order to integrate more perturbative contributions.
Because of the infrared safety, the uncertainty due to the
choice of $Q_0^2$ is minimal.

Substituting (\ref{eq:ct}) into (\ref{eq:fn}) and replacing
$4/r^2$ by $Q'^2$, one
recovers the familiar integrated form of the GLAP 
evolution equation at small $x$
\begin{eqnarray}
f_N(x,Q^2) & = & f_N(x,Q_0^2)+\frac{1}{4\pi}\int_x^1\frac{dx'}{x'}\nonumber \\
& & \times
\int_{Q_0^2}^{Q^2} \frac{dQ'^2}{Q'^2}
\alpha_s(Q'^2)g_N^{\rm DLA}(x',Q'^2)\; .\label{eq:fnap}
\end{eqnarray}
Note that (\ref{eq:fnap}) is not an integral equation since it does not
involve the sea-quark contribution on the right hand side. It is
derived in the DLA where the small-$x$ behavior is driven by the
gluon contribution. To improve the larger $x$ behavior,
we insert a splitting function $P_{gq}(z)=z^2+(1-z)^2$ where
$z=x/x'$ in (\ref{eq:fnap}) and add the valence quark contributions
to $F_2^N$
\begin{eqnarray}
F_2^N(x,Q^2)=\sum_ie_i^2[xf_N(x,Q^2)+x\upsilon_i(x,Q^2]\; ,
\end{eqnarray}
where $x\upsilon$'s are the valence quark distributions,
which are given by the GRV \cite{mrs} parameterization (there is
a very little difference among different parameterizations 
for valence quarks).

\subsection{Scaling Violation of $F_2^N(x,Q^2)$ and Gluon Density} 
The $Q^2$ evolution or the scaling violation of the nucleon gluon density
in the small-$x$ region is quite different from that in the large-$x$
region in the context of GLAP equation. In the region of
small $x$, one has to
sum up large $\alpha_s\ln 1/x$ contributions in addition to 
$\alpha_s\ln Q^2$ terms represented
by the ladder diagrams with strong scale ordering in the leading
logarithm approximation (LLA). To avoid summing up two large
logarithms, one considers the DLA, i.e., the
most leading diagrams involving $(\alpha_s\ln Q^2\ln 1/x)^n$
at each order $n$. Such a double
logarithm appears only when vector particles (gluons) propagate in the
$t$-channel of the ladder diagram as argued by Gribov, Levin and
Ryskin in \cite{glr}. Thus, the scale dependence of the nucleon
quark distribution 
is directly related the small-$x$ behavior of the gluon distribution.
 
In general, the DLA is valid in the kinematical region where
$\alpha_s\ln 1/x\ln Q^2/Q_0^2\sim 1$ while $\alpha_s\ln 1/x \ll 1$ and
$\alpha_s\ln Q^2/Q_0^2 \ll 1$ as well as $\alpha_s\ll 1$. 
The precise validity range of DLA or of GLAP equation is currently
still under debate. The general belief is that it describes
a region up to a moderately large $Q^2$ ($\sim 20$ GeV$^2$) and a 
moderately small $x$, $10^{-5}<x<10^{-2}$. Formally, 
in the small-$x$ region, one should sum the large 
$\alpha_s\ln 1/x $ terms using a BFKL resummation technique \cite{bfkl}.
However, it has been shown recently by Ball and Forte \cite{dla}
that a gluon density obtained in the DLA together with the
resummed leading logarithm in $Q^2$ gives a very good description
of the double-asymptotic scaling observed at HERA \cite{h1}.
The unitarized gluon density in the DLA has a simple approximate
form \cite{glr,dla},
\begin{eqnarray}
xg_N^{\rm DLA}=xg_N(x,Q_0^2)I_0(y)\; , \label{eq:dla}
\end{eqnarray}
where $I_0$ is the Bessel function and 
\begin{eqnarray}
y=2c\sqrt{\ln \left (\frac{1}{x}\right )
\ln \left (\frac{t}{t_0}\right )}\; ,\label{eq:y}
\end{eqnarray}
with $t\equiv \ln Q^2/\Lambda^2$ and $c=\sqrt{36/25}$,
provided that input $xg_N(x,Q_0^2)$ is not singular in $x$. 
The asymptotic behavior of 
$I_0(y)$ is that $I_0(y)\rightarrow 1$ as $y\rightarrow 0$ and
$I_0(y)\rightarrow e^y/\sqrt{2\pi y}$ as $y\rightarrow \infty$.
That is,
in this DLA, $xg_N$ increases faster than any power of $\ln 1/x$,
but slower than a power of $1/x$. To improve the DLA at a large
$Q^2$ ($>20$ GeV$^2$), the resummation of single logarithm, 
$\alpha_s\ln Q^2/Q_0^2$, is necessary. 
This results in an additional
$Q^2$ dependence of the gluon distribution,
a multiplicative factor $(t/t_0)^{-\delta}$ in (\ref{eq:dla})
where $\delta =61/45$ \cite{dla}.
We find that this modification has very little numerical effect
on the nuclear shadowing ratio in the region of interest.

In a semiclassical form $xg_N=x^{-\omega}(Q^2)^\gamma$.
One defines the exponents $\gamma =\partial \ln xg_N/\partial t$ and
$\omega =\partial xg_N/\partial \ln (1/x)$. In the DLA,
\begin{eqnarray}
\gamma =\frac{\ln 1/x}{t\ln t/t_0}\omega\; .\label{eq:exp}
\end{eqnarray}
The anomalous dimension can be large in the small-$x$ and -$t$
region. As $t\rightarrow t_0$, $\omega\rightarrow 0$ and $\gamma$
is finite.
It is important to emphasize that the DLA form (\ref{eq:dla}) 
is only valid when $Q^2\gg Q_0^2$.  

Our goal is to pin down the gluon density in the nucleon. In our
approach, the slope of $F_2^N$ in $Q^2$,
${d F_2^N}/{d\ln Q^2}$, is virtually a direct 
indicator of the magnitude of the gluon distribution. We require
that the gluon distribution in the
DLA as given in (\ref{eq:dla}) describes reasonably well 
the scaling violation of $F_2^N$ for $Q^2$ ranging from $0.85$ GeV$^2$
to $100$ GeV$^2$ where the experimental data exist in the small-$x$ 
region. As it is turned out, this almost uniquely determines the 
gluon distribution.

\begin{figure}
\vskip -0.25in
\centerline{\epsfig{figure=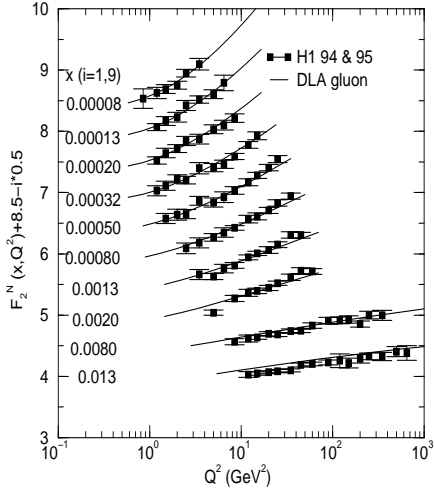,width=3in,height=3in}}
\vskip -0.2in
\caption{The $Q^2$-dependence of the
nucleon structure function calculated using
the DLA and compared with experiment data [24].}
\label{fig:h1_q}
\end{figure}

\begin{figure}
\centerline{\epsfig{figure=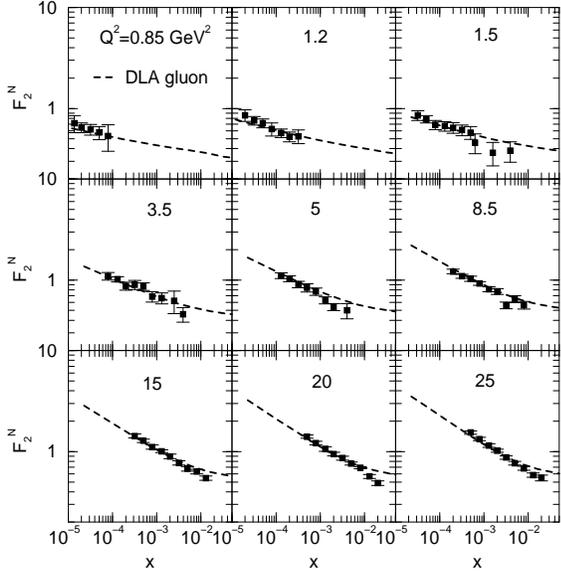,width=3.5in,height=3in}}
\caption{The $x$-dependence of the
nucleon structure function calculated using
the DLA and compared with the experiment data.}
\label{fig:h1}
\end{figure}

There are two unknown quantities in (\ref{eq:dla}): $Q_0^2$ and the initial
condition $xg_N(x,Q_0^2)$. $Q_0^2$ determines where the perturbative
calculation should begin to be valid and where
the non-perturbative contribution
can be separated into an initial condition $xg_N(x,Q_0^2)$. We would 
like to have a reliable perturbative description of the structure
function in the range from $1$ to a few GeV which is 
extremely important in understanding the evolution of nuclear shadowing.
This leads to choosing $Q_0^2$ to be rather small so that the DLA sets in
at a relatively low $Q^2=1\sim 2$ GeV$^2$. 
Since $xg_N(x,Q_0^2)$ has to be non-singular in the small $x$
regime, we take $xg_N(x,Q_0^2)=c_gxg_N^{\rm GRV}(x,Q_0^2)$ at 
an initial scale $Q_0^2$ since the initial gluon distribution 
in GRV is fairly flat while it has the proper behavior at $x\rightarrow 1$. 
The initial
condition for $F_2^N$ is taken to be in the form motivated by a soft pomeron
with an intercept $\alpha =1+0.08$ \cite{landshoff}
\begin{eqnarray}
F_2^N(x,Q_0^2)=c_Px^{-0.08}(1-x)^\eta\; ,\label{eq:f2n0}
\end{eqnarray}
where $\eta =10$ accounts for the behavior as $x\rightarrow 1$.
We determine $Q_0^2$, $c_g$ and $c_P$ by requiring that the calculated
$Q^2$-dependence of $F_2^N$ at $x=8\times 10^{-5}$ fits the recent
measurement from H1 Collaboration \cite{h1}. We find the following 
unique solution
\begin{eqnarray}
Q_0^2=0.4 \; ; \quad c_g=2.2\; ; \quad c_P=0.2\; .
\end{eqnarray}
The $Q^2$-dependence of $F_2^N$ at other values of small $x$ are 
then predicted using (\ref{eq:fnap}). 
In Fig.\ \ref{fig:h1_q}, we show our results for $F_2^N$ 
compared to the experimental
values \cite{h1}. The gluon density in (\ref{eq:dla}) describes the
scale dependence of $F_2^N$ very well starting at a relatively low 
$Q^2$, $Q^2\sim 0.85$ GeV$^2$.   
The $x$-dependence of the nucleon structure function can also be
predicted from (\ref{eq:fnap}).
Fig.\ \ref{fig:h1}
shows the calculated $F_2^N(x,Q^2)$ from (\ref{eq:fnap})
as function of $x$ at various $Q^2$ as compared to H1 measured values
\cite{h1}. The agreement is evident. Starting with a fairly
flat distribution $F_2^N(x,Q_0^2)$, the nucleon structure function
develops a more and more singular shape in $x$ from the perturbative 
evolution governed by the gluon distribution. 
Some deviation is visible for $x>10^{-2}$, a region where
we do not expect DLA  to be valid.
The results shown in Fig.\ \ref{fig:h1_q} and Fig.\ \ref{fig:h1} 
demonstrate that 
the gluon density in the DLA given in (\ref{eq:dla}) faithfully captures
the essential feature of $Q^2$- and $x$-evolution of the nucleon
structure function.
The fact that we can describe
$F_2^N$ even at very low $Q^2$, $0.85<Q^2<5$ GeV$^2$
indicates that the perturbative approach is reliable in the semihard region.

\subsection{Gluon Distribution in the DLA}
Similarly to the quark case, one can consider the
virtual $gg$ pair interacting with the
nucleons in the nuclear target.
Using the well-known relationship between the gluon density and the
deeply-inelastic cross section \cite{mueller},
the gluon density can be written as \cite{agl}
\begin{eqnarray}
xg_N(x,Q^2)=\frac{2}{\pi^4}\int_0^1dz\int d^2{\bf r}
|\psi(z,{\bf r})|^2\sigma_{ggN}({\bf r})\; , \label{eq:gsig}
\end{eqnarray}
where the wave function squared for a gluon pair is 
\begin{eqnarray}
|\psi(z,{\bf r})|^2 & = & \frac{1}{z(1-z)}\left [(a^2K_2(ar)-aK_1(ar)/r)^2
\right .\nonumber \\
& & \left .+a^2K_1(ar)^2/r^2\right ]
\stackrel{ar\ll 1}{\longrightarrow}  \frac{2}{zr^4}\; ,\label{eq:gwf}
\end{eqnarray}
and the cross section in the DLA is \cite{strikman}
\begin{eqnarray}
\sigma_{ggN}({\bf r})=\frac{3\pi^2}{4}
r^2\alpha_s(4/r^2)x'g_N^{\rm DLA}(x',4/r^2)
\; . \label{eq:gct}
\end{eqnarray}
Note $\sigma_{ggN}=9\sigma_{q\bar qN}/4$. 
Substituting (\ref{eq:gwf}) and (\ref{eq:gct}) into 
(\ref{eq:gsig}), one obtains
\begin{eqnarray}
g_N(x,Q^2)& = & g_N(x,Q_0^2)+\frac{1}{2\pi}\int_x^1 \frac{dx'}{x'}
\int_{Q_0^2}^{Q^2}\frac{dQ'^2}{Q'^2}\nonumber\\
& & \times\alpha_s(Q'^2)\left (6\frac{x'}{x}\right )g_N^{\rm DLA}(x',Q'^2)\; ,
\label{eq:gap}
\end{eqnarray}
a form in which the small-$x$ limit splitting function in the DLA
can be easily identified: $P_{gg}(z)\simeq 6/z$ where $z=x/x'$.
Comparing (\ref{eq:gap}) with (\ref{eq:fnap}), one finds that due
to different splitting functions, the gluon density increases as
$Q^2$ increases $12$ times faster than 
the quark density, leading to a much stronger scaling violation.
Similar to the quark case, (\ref{eq:gap}) is not to be interpreted
as the integral equation since the right hand side does not contain
$g_N$ but $g_N^{\rm DLA}$.
Namely, $g_N^{\rm DLA}$ is the
gluon density seen by a $gg$ pair with a frozen transverse separation
while $g_N$ is the gluon density seen by the probe, e.g., the virtual
Higgs boson. However, since the integration kernel in (\ref{eq:gap})
coincides with that of the DLA evolution equation, $g_N$ should be
very close to $g_N^{\rm DLA}$. We confirm that numerically they differ
only by few percent. 

\section{Glauber Multiple-Scattering Model}
\label{sec:glb}
The Glauber multiple-scattering theory \cite{glauber}
treats a nuclear collision
as a succession of collisions of the probe
with individual nucleons within nucleus.
A partonic system ($h$), being a $q\bar q$ or a $gg$ fluctuation, 
can scatter coherently from several or all nucleons during its passage
through the target nucleus. The interference between the multiple scattering
amplitudes causes a reduction of the $hA$ cross section compared
to the naive scaling result of $A$ times the respective $hN$ cross 
section, the origin of the nuclear shadowing. In the high-energy 
eikonal limit, the phase shift of the elastic $hA$ scattering
amplitude is just the sum of phase shifts of individual $hN$ scatterings.
The total $hA$ cross section is determined by the imaginary part
of the forward elastic scattering amplitude as governed by the 
unitarity relation or the optical theorem. The unitarity of 
the total cross section is thus built-in within the Glauber theory. 

In the high-energy eikonal limit where the projectile's momentum
${\bf k}$ is much larger than the inverse of the interaction
radius, one describes the scattering amplitude by the Fourier
transformation of the so-called profile function $\Gamma ({\bf b})$
in the impact parameter space
\begin{eqnarray}
f({\bf k},{\bf k'})=\frac{ik}{2\pi}\int d^2{\bf b}
e^{i({\bf k}-{\bf k'})\cdot {\bf b}}\Gamma ({\bf b})\; ,\label{eq:amp}
\end{eqnarray}
where ${\bf k'}$ is the outgoing wave momentum. In the high energy
scattering process, the collision is almost completely absorptive
corresponding to an approximately real $\Gamma ({\bf b})$ and 
pure imaginary amplitude. The total cross section is given by the
optical theorem
\begin{eqnarray}
\sigma_{\rm tot}=\frac{4\pi}{k}{\rm Im}f(0)=
2\int d^2{\bf b}\Gamma ({\bf b})\; . 
\end{eqnarray}
To extend these results to nuclei, one assumes that each collision
with the nucleon modifies the piece of the incident wave passing through
it by a factor $1-\Gamma_N ({\bf b}-{\bf s})$ where ${\bf s}$ is the
transverse position of the nucleon. The overall modification of 
the wave is thus given by
\begin{eqnarray}
\prod_{i=1}^A\left [1-\Gamma_N ({\bf b}-{\bf s}_i)\right ]\; . \label{eq:mod}
\end{eqnarray}
Clearly, only those nucleons in the vicinity of impact parameter ${\bf b}$
can influence $\Gamma_A$.
The nucleus profile function is just the factor in (\ref{eq:mod}) 
multiplied by the probability of finding nucleons in the positions
${\bf r}_1,...,{\bf r}_A$
\begin{eqnarray}
|\psi ({\bf r}_1,...,{\bf r}_A)|^2=\prod_{i=1}^A \rho ({\bf r}_i)\; ,
\end{eqnarray}
where $\rho ({\bf r}_i)d^3{\bf r}_i$ is the probability of finding 
a nucleon at ${\bf r}_i$. The nucleon number density is then
$\rho_A({\bf r})=\sum_i \rho ({\bf r}_i)$. For large $A$, 
the nucleus profile is calculated
\begin{eqnarray}
\Gamma_A({\bf b}) & = & 1-\prod_{i=1}^A \int d^3{\bf r}_i \rho ({\bf r}_i)
[1-\Gamma_N ({\bf b}-{\bf s}_i)]\nonumber\\
& = & 1-\prod_{i=1}^A[1-\int d^3{\bf r}_i\rho ({\bf r}_i)
\Gamma_N ({\bf b}-{\bf s}_i)]\nonumber\\
& = & 1-\left [1-\frac{1}{A}\int d^2{\bf s}dz\rho_A ({\bf s},z)
\Gamma_N ({\bf b}-{\bf s})\right ]^A \label{eq:eik}\\
&\simeq & 1-\exp \left [-\int d^2{\bf s}T_A({\bf s})
\Gamma_N ({\bf b}-{\bf s})\right ]\nonumber \; , 
\end{eqnarray}
where we have used $A\gg 1$. We shall use a Gaussian parameterization
of the nucleus thickness function defined by 
\begin{eqnarray}
T_A({\bf b})\equiv \int dz \rho_A ({\bf b},z)=
\frac{A}{\pi R_A^2}e^{-b^2/R_A^2}\; ,
\end{eqnarray}
where the mean nucleus radius $R_A^2=0.4R_{WS}^2$ and 
$R_{WS}=1.3A^{1/3}$ fm.

Since the nucleon has a smaller transverse size than the typical nuclear
dimension, one approximates the nucleon profile function in (\ref{eq:eik})
by $\Gamma_N ({\bf b}-{\bf s})\simeq \sigma_{hN}\delta^2({\bf b}-{\bf s})/2$
and obtains the total $hA$ cross section
\begin{eqnarray}
\sigma_{hA}=\int d^2{\bf b} 2
\left [ 1-e^{-\sigma_{hN}T_A({\bf b})/2}\right ] \; .\label{eq:glb}
\end{eqnarray}
For a Gaussian parametrization of $T_A({\bf b})$, the integration on the
impact parameter can be done exactly leading to
\begin{eqnarray}
\sigma_{hA} & = & 2\pi R_A^2\left [\kappa_h -\frac{1}{4}\kappa_h^2+\cdots
+ (-1)^{n-1}
\frac{\kappa_h^n}{nn!}+\cdots\right ] \nonumber\\
& = & 2\pi R_A^2\left [ \gamma +\ln \kappa_h +E_1(\kappa_h )\right ]\; ,
\end{eqnarray}
where $\kappa_h =A\sigma_{hN}/(2\pi R_A^2)$, $\gamma =0.5772$ is 
Euler's constant and $E_n(x)$ is the exponential integral 
function of rank $n$. The dimensionless quantity $\kappa_h$ serves as
an impact parameter averaged effective number of scatterings.
For small value of $\kappa_h$, 
$\sigma_{hA}\rightarrow 2\pi R_A^2\kappa_h =A\sigma_{hN}$, the total $hA$
cross section is proportional to $A$ or the nuclear volume. In the 
limit $\kappa_h \rightarrow \infty$, the destructive interference between
multiple scattering amplitudes reduces the cross section to
$\sigma_{hA}\rightarrow 2\pi R_A^2 (\gamma +\ln \kappa_h )$. Namely,
the effective number of scatterings is large and 
the total cross section approaches the geometric limit $2\pi R_A^2$, a
surface term which varies roughly as $A^{2/3}$. In the kinematical region
that we are interested in none of these limits are actually approached. 
The shadowing effect is determined by the strength of $\kappa_h$, which
is different for a $h=q\bar q$ than for a $h=gg$ case.

So far we have sketched the necessary assumptions for applying the
Glauber theory for the elastic or the quasielastic scattering, that is,
the longitudinal momentum remains the same during the passage of
$h$ through the nucleus. This condition can be relaxed to the forward
production processes where there is a longitudinal momentum transfer 
appropriate to the $q\bar q$ or the $gg$ pair production. In this case,
the scattering amplitude in (\ref{eq:amp}) is modified to include the
integration of the longitudinal position $z$ due to the fact that an
incident wave $e^{ik_z^\gamma z}$ for the photon $\gamma$ is changed to
a final wave $-\Gamma ({\bf b})e^{ik_z^hz}$ for particle $h$. In general,
$k_z^\gamma \neq k_z^h$, and there is a phase difference between the $\gamma$
and $h$ waves behind the target, i.e., $e^{i\Delta k_zz}$ where 
$\Delta k_z=k_z^\gamma -k_z^h$. 
This leads to  a coherence length cutoff which is governed by
$l_c\simeq 1/\Delta k_z$. If $l_c>R_A$, the hadronic fluctuation $h$ interacts
coherently with all nucleons within the nucleus and the Glauber theory
is valid. On the other hand, if $l_c<R_A$, $h$ interacts coherently only with
some of the nucleons and the interference effect is reduced.

To estimate the coherence length $l_c$, one calculates the change of the
longitudinal momenta in the small-$x$ region
\begin{eqnarray}
\Delta k_z & = &\sqrt{\nu^2+Q^2}-\sqrt{\nu^2-M^2} \nonumber\\
& \simeq & \frac{Q^2+M^2}{2\nu}
=(x+x')m\; ,
\end{eqnarray}
where $M$ is the invariant mass of the  $q\bar q$ or the $gg$ pair
as defined in (\ref{eq:xp}). Since the cross section is dominated by 
$x'$ near $x$, the coherence length $l_c\simeq 1/(2xm)$ exceeds 
the radius of heavy nuclei $R_A\sim 4$ fm as long as $x<2\times 10^{-2}$ 
and the Glauber multiple scattering model is adequate. 
For $x>2\times 10^{-2}$, the Glauber approximation breaks down and one
should take into account the finite coherence length. 
Although we are mainly concerned with the small-$x$ behavior of the
nuclear shadowing, we would like to get some handle on the coherence
length cutoff in the moderately large-$x$ region.
We propose to take a probability approach \cite{agl,boris}
and introduce a coherence length form
factor $\tau (\Delta k_z) =\exp [-R_A^2(\Delta k_z)^2/4]$ \cite{agl} while
preserving the nice structure of the Glauber formula.
The probability for a double scattering
is reduced by $\tau$, and that for a $n$-scattering is reduced
by $\tau^{n-1}$ so that one may make the following replacement
in (\ref{eq:glb})
\begin{eqnarray}
2\left [ 1-e^{-\sigma_{hN}T_A({\bf b})/2}\right ] \rightarrow
\frac{2}{\tau }
\left [ 1-e^{-\sigma_{hN}T_A({\bf b})\tau /2}\right ]\; .
\label{eq:tau}
\end{eqnarray}
As $\tau \rightarrow 1$, the original Glauber formula is preserved, while
as $\tau \rightarrow 0$, (\ref{eq:tau}) gives $\sigma_{hA}=A\sigma_{hN}$,
i.e., there is no shadowing. In the kinematical region $x<10^{-2}$,
the coherence length cutoff does not play a role. Even for $x>10^{-2}$,
because the cross section is already very small, 
the effect is not very large. 

\section{Nuclear Quark Distribution $F_2^A$}
\label{sec:fa}
We now examine the effect of the gluon density in (\ref{eq:dla})
on the shadowing of the nuclear structure function.
Replacing
$\sigma_{hN}$ in (\ref{eq:fn}) and (\ref{eq:gsig}) by $\sigma_{hA}$
as calculated by the Glauber multiple scattering model, one obtains
for (sea)quarks
\begin{eqnarray}
xf_A(x,Q^2) & = & xf_A(x,Q_0^2)+\frac{3R_A^2}{8\pi^2}x\int_x^1 \frac{dx'}{x'^2}
\int_{Q_0^2}^{Q^2}
dQ'^2\nonumber\\
& & \times\frac{1}{\tau}\left [ \gamma +\ln (\kappa_q\tau ) +E_1(\kappa_q\tau )
\right ]\; ,\label{eq:fa}
\end{eqnarray}
where the effective number of scatterings $\kappa$ is given by
\begin{eqnarray}
\kappa_q(x,Q^2) =\frac{2A\pi}{3R_A^2Q^2}\alpha_s(Q^2)xg_N^{\rm DLA}(x,Q^2)
\; , \label{eq:kappa}
\end{eqnarray}
Expanding $[\cdots ]$ in (\ref{eq:fa}) and keeping
only the first term which is proportional to $\kappa_q$, one recovers
the results for the nucleon in the same DLA approximation.

\begin{figure}
\centerline{\epsfig{figure=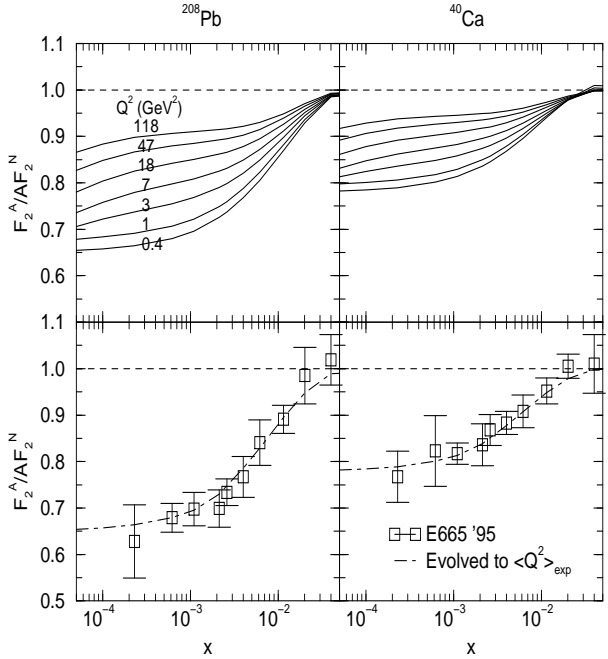,width=3.5in,height=3.5in}}
\caption{The $x$-dependence of nuclear quark shadowing ratio
$R_q=F_2^A/(AF_2^N)$ at various $Q^2$ for $^{208}$Pb and $^{40}$Ca.
The lower panels represent the parameterizations of initial 
shadowing ratios evolved to the experimental 
$\langle Q^2\rangle_{\rm exp}$ values. The non-saturation
of $R_q(x)$ as $x$ decreases at large $Q^2$ is the feature of the
perturbative shadowing mechanism.}
\label{fig:rq}
\end{figure}

The nuclear structure function $F_2^A(x,Q^2)$ is defined through the
total $\gamma^*$-$A$ cross section, similar to $F_2^N$ in (\ref{eq:vir}).
To improve the large-$x$ behavior, we add to $F_2^A$ the valence
quark contributions and assume that there is no shadowing for the valence
quarks
\begin{eqnarray}
F_2^A(x,Q^2)=\sum_ie_i^2[xf_A(x,Q^2)+Ax\upsilon_i(x,Q^2]\; .\label{eq:f2a}
\end{eqnarray}
In principle, the initial condition $xf_A(x,Q_0^2)$ in (\ref{eq:fa}) 
may be calculated in the vector-meson-dominance (VMD) model. Fortunately,
there have been some experimental measurements on the nuclear quark
shadowing in the low-$Q^2$ region so that the initial condition can be
expressed in terms of the production of the initial nucleon structure
function and the shadowing ratio $R_0^q(x)$
\begin{eqnarray}
F_2^A(x,Q_0^2)=AR_0^q(x)F_2^N(x,Q_0^2)\; , \label{eq:rq0}
\end{eqnarray}
where $F_2^N(x,Q_0^2)$ is given in (\ref{eq:f2n0}). 
Recently E665 Collaboration has measured
the nuclear shadowing effect
in muon scatterings off 
$^{40}$Ca and $^{208}$Pb at low $x$ \cite{fermilab}. In general,
the small $x$ data are measured at low $Q^2$, e.g., for $x=6\times 10^{-4}$
the average $\langle Q^2\rangle =0.26$ GeV$^2$ while $x=6\times 10^{-3}$
corresponds to $\langle Q^2\rangle =1.35$ GeV$^2$. We parametrize the 
initial shadowing ratio $R_0^q(x)$ at $Q_0^2=0.4$ GeV$^2$ and use 
(\ref{eq:f2a}) to calculate the nuclear structure function at the measured
$\langle Q^2\rangle$ values at different $x$ values. The nuclear shadowing
ratio is defined by
\begin{eqnarray}
R_q(x,Q^2) & = & \frac{F_2^A(x,Q^2)}{AF_2^N(x,Q^2)} \nonumber\\
& = &
\frac{R^0_q(x)F_2^N(x,Q_0^2)+\Delta F_2^A(x;Q^2,Q_0^2)}{F_2^N(x,Q_0^2)+
\Delta F_2^N(x;Q^2,Q_0^2)}
\; ,\label{eq:rq}
\end{eqnarray}
where $\Delta F_2\equiv F_2-F_2(Q_0^2)$ 
can be explicitly calculated by performing
the double integration over $x'$ and $Q'^2$ in (\ref{eq:fa}).
The results for $^{40}$Ca and $^{208}$Pb are shown in
Fig.\ \ref{fig:rq}. 

\begin{figure}
\vskip -0.2in
\centerline{\epsfig{figure=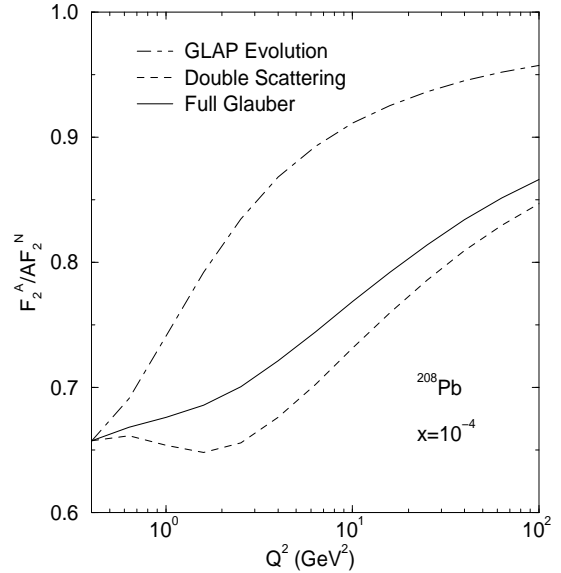,width=4in,height=3.5in}}
\vskip -0.2in
\caption{The $Q^2$-dependence of nuclear quark shadowing ratio
$R_q=F_2^A/(AF_2^N)$ at $x=10^{-4}$ for $^{208}$Pb.} 
\label{fig:qrq}
\end{figure}

The lower panels of Fig.\ \ref{fig:rq} show our parametrizations of initial
shadowing ratios at $Q_0^2$ evolved to the measured $\langle Q^2\rangle$
compared to the experimental data from E665. For $\langle Q^2\rangle <Q_0^2$,
we take it to be at the minimal scale $Q_0^2$. As is evident from 
the comparison, our parameterizations faithfully represent the experimental
data at small $x$. The shadowing ratios $R_q(x,Q^2)$ at high values
of $Q^2$ are shown in the upper panels of Fig.\ \ref{fig:rq}. 
We note that at large $Q^2$ the nuclear
shadowing effect is reduced but does not diminish, i.e.\ 
it is not a higher twist effect. This can be understood as the interplay
between the perturbative and the non-perturbative shadowing mechanisms. 
The nuclear quark density is initially shadowed by $R^0_q$ at $Q_0^2$ 
compared to the free nucleon case due to some non-perturbative mechanism
such as coherent scatterings of the $\rho$-meson as discussed
in the VMD model \cite{vmd}. As $Q^2$ increases from $Q_0^2$, the nucleon
structure function increases by $\Delta F_2^N(x;Q^2,Q_0^2)$ as governed
by the GLAP evolution equation
in the DLA. If the nuclear $F_2^A$ increases by the
same amount (times $A$), i.e., we neglect the multiple scattering effect
and evolve it according to the same GLAP equation, the initial
shadowing effect in (\ref{eq:rq}) decreases quickly due to the strong
scaling violation at low $Q^2$ ($\Delta F_2^N(x;Q^2,Q_0^2)$ is not small
compared to $F_2^N(x,Q_0^2)$). This is illustrated in Fig.\ \ref{fig:qrq}
where the shadowing ratio is calculated by
taking the first term in the expansion $[\cdots ]$ in
(\ref{eq:fa}) and is  plotted as a function of $Q^2$ at 
$x=10^{-4}$. The shadowing effect disappears quickly when
both nucleon and nuclear structure functions evolve according 
to the GLAP equation.

However, because $xg_N^{\rm DLA}$ increases
rapidly as $Q^2$ increases, the growth of the gluon density in a nucleon
compensates for the suppression due to the smallness of the size of
the $q\bar q$ pair.
As a result, $\kappa_q$ is not small in the semihard $Q^2$ region. In fact,
$\kappa_q$ increases as $Q^2$ increases from $Q_0^2$ to 2 GeV$^2$ 
and becomes greater than $1$. The perturbative rescattering effect becomes 
important and needs to be taken into account in the Glauber formalism.
The effect of  multiple scatterings slows down the $Q^2$ evolution of $F_2^A$
as compared to $F_2^N$, i.e., 
\begin{eqnarray}
\Delta F_2^A(x;Q^2,Q_0^2)<\Delta F_2^N(x;Q^2,Q_0^2)\; ,
\end{eqnarray}
which provides a perturbative shadowing mechanism beyond $Q_0^2$.
This is illustrated in Fig.\ \ref{fig:qrq} where the $Q^2$ evolution 
is indeed slower than the GLAP evolution case. As $Q^2$ becomes even
larger ($Q^2>5$ GeV$^2$), $\kappa_q$ decreases fast and the 
probability for rescattering becomes
 small. In this case, both $F_2^A$ and $F_2^N$ evolve
slowly and so does the shadowing ratio. Also plotted in Fig.\ \ref{fig:qrq}
is the double scattering case where we take the first two terms
in the expansion
\begin{eqnarray}
\left [ \gamma +\ln \kappa_q+E_1(\kappa_q)\right ]\simeq\kappa_q -
\frac{1}{4} \kappa_q^2\; .
\end{eqnarray}  
As it is evident from Fig.\ \ref{fig:qrq}, the double scattering overestimates 
the amount of shadowing. Nevertheless, it
accounts for most of the multiple scattering effect.
The interplay between soft physics and the partonic picture of the
nuclear shadowing is also evident in the $x$-dependence of $R_q$
in Fig.\ \ref{fig:rq}. The apparent flatness of the 
shadowing ratio at low $Q^2$ in the small-$x$ region is 
altered by the perturbative
evolution. This is due to the singular behavior of $xg_N^{\rm DLA}$
as $x\rightarrow 0$ at large $Q^2$ leading to the strong 
$x$-dependence of the
effective number of scatterings, $\kappa_q(x)$. Thus, the saturation
of the $R_q$ as $x$ decreases is a feature only at low $Q^2$ ($<2$ GeV$^2$) 
where the gluon density is not so singular, and we do not expect 
saturation to happen for higher values of $Q^2$.

\section{Predicting Gluon Shadowing}
\label{sec:ga}  
In the previous section, we have emphasized the importance of 
the interplay between the non-perturbative 
and perturbative shadowing mechanisms in order to predict the 
nuclear quark shadowing at large $Q^2$. This would suggest that 
the prediction for the  nuclear gluon shadowing is uncertain because
there is very little experimental information on the gluon shadowing
at low $Q^2$. However, as we shall demonstrate below, this is not
the case and the gluon shadowing can be unambiguously predicted for
large $Q^2$ in the context of perturbative QCD.

The nuclear gluon distribution can be calculated in the
Glauber model similarly to the quark case \cite{agl}
\begin{eqnarray}
xg_A(x,Q^2) & = & xg_A(x,Q_0^2)+\frac{2R_A^2}{\pi^2}\int_x^1\frac{dx'}{x'}
\int_{Q_0^2}^{Q^2}
dQ'^2\nonumber\\
& & \times\frac{1}{\tau}\left [ \gamma +\ln (\kappa_g\tau ) +E_1(\kappa_g\tau )
\right ]\; ,\label{eq:ga}
\end{eqnarray}
The main difference between the quark and gluon cases is 
different wave functions of the virtual probe and the
effective number of scatterings,  $\kappa_g=9\kappa_q/4$ due to different
color representations that the quark and the gluon belong to. 
As a result, 
the gluon density increases as $Q^2$ increases $12$ times faster 
than the sea-quark density at small $x$.
The two important effects which make gluon shadowing quite different
from quark shadowing are a stronger scaling violation in the semihard
scale region and a larger perturbative shadowing effect. 
This can be seen by considering the shadowing ratio
\begin{eqnarray}
R_g(x,Q^2) & = & \frac{xg_A(x,Q^2)}{Axg_N(x,Q^2)}\label{eq:rg}\\
& = & \frac{xg_N(x,Q_0^2)R_g^0(x)+
\Delta xg_A(x;Q^2,Q_0^2)}{xg_N(x,Q_0^2)+\Delta xg_N(x;Q^2,Q_0^2)}
\; , \nonumber
\end{eqnarray}
where $R_g^0(x)$ is the initial shadowing ratio at $Q_0^2$ and
$\Delta xg(x;Q^2,Q_0^2)$ is the change of the gluon distribution
as the scale changes from $Q_0^2$ to $Q^2$. The strong scaling 
violation due to a larger $\kappa_g$ at small $x$ causes 
$\Delta xg_N(x;Q^2,Q_0^2)\gg xg_N(x,Q_0^2)$ as $Q^2$ becomes greater than
$1\sim 2$ GeV$^2$. The dependence of $R_g(x,Q^2)$ on the initial
condition $R_g^0(x)$ diminishes and the perturbative shadowing
mechanism takes over, i.e.
\begin{eqnarray}
 R_g(x,Q^2)\rightarrow \frac{\Delta xg_A(x;Q^2,Q_0^2)}{\Delta 
xg_N(x;Q^2,Q_0^2)}
\; , 
\end{eqnarray}
for $Q^2>2$ GeV$^2$.

The above statement can be made more rigorous by taking a partial
derivative of (\ref{eq:rg}) with respect to $\ln Q^2$ at fixed $x$
\begin{eqnarray}
\frac{\partial R_g(x,Q^2)}{\partial \ln Q^2}=\gamma (x,Q^2)\left [
R_g^{\rm PT}(x,Q^2)\right.\nonumber \\
\left. -R_g (x,Q^2)\right ] \;,\label{eq:rgq}
\end{eqnarray}
where the anomalous dimension $\gamma (x,Q^2)$ and the 
so-called perturbative shadowing ratio $R_g^{\rm PT}$ are defined
\begin{eqnarray}
\gamma (x,Q^2) & = &\frac{\partial \ln xg_N(x,Q^2)}{\partial \ln Q^2}
\nonumber\\
R_g^{\rm PT}(x,Q^2) & = &\frac{\partial xg_A(x,Q^2)/\partial 
\ln Q^2}{A\partial xg_N(x,Q^2)/\partial\ln Q^2}\; .\label{eq:rpt}
\end{eqnarray}
Note that $R_g^{\rm PT}$ is the ratio of the derivative of $xg_A$
to that of $xg_N$, which is independent of the initial shadowing
ratio and can be calculated perturbatively. In the small-$x$ 
region
\begin{eqnarray}
R_g^{\rm PT}\simeq \frac{\gamma +\ln \kappa_g +E_1(\kappa_g )}{\kappa_g}
=1-\frac{\kappa_g}{4} +\frac{\kappa_g^2}{18}-\cdots\; .\label{eq:rptk}
\end{eqnarray}

The interesting feature of (\ref{eq:rgq}) is the negative sign in 
front of $R_g$ on the right hand side such that it is
self-stabilizing. If $R_g<R_g^{\rm PT}$ at $Q^2$ and 
${\partial R_g}/{\partial \ln Q^2}>0$, the shadowing ratio  
increases with $Q^2$; while for $R_g<R_g^{\rm PT}$, the shadowing ratio
decreases with $Q^2$. As a result, the $Q^2$-evolution of $R_g$ is 
such that it is always in the process of approaching $R_g^{\rm PT}$.
Because $R_g^{\rm PT}$ also changes with $Q^2$, the $Q^2$-evolution of $R_g$
never stops. The anomalous dimension $\gamma (x,Q^2)$ determines how
fast $R_g$ is approaching $R_g^{\rm PT}$. 
\begin{figure}
\vskip -0.3in
\centerline{\epsfig{figure=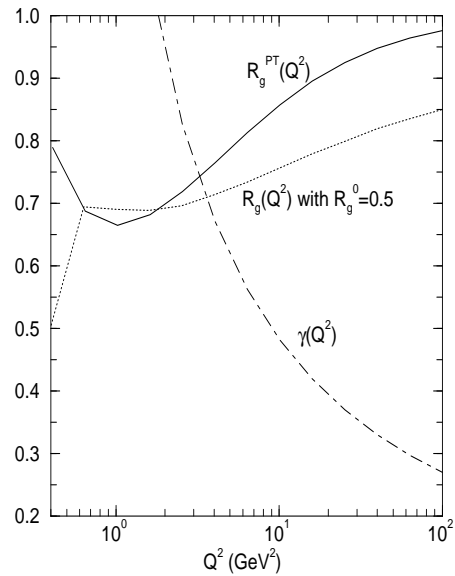,width=3.5in,height=3.5in}}
\vskip -0.2in
\caption{The $Q^2$-dependence of the anomalous dimension
$\gamma (x,Q^2)$ and the perturbative shadowing ratio $R_g^{\rm PT}(x,Q^2)$
at $x=10^{-4}$ for $^{208}$Pb. Also plotted is the shadowing ratio
$R_g$ with an initial condition $R_g^0=0.5$.} 
\label{fig:rgpt}
\end{figure}
Introduce an important
scale $Q^2_{\rm SH}=3$ GeV$^2$ (``SH'' stands
for ``semihard'') where $\gamma (x,Q^2_{\rm SH})\simeq $O(1)
for small values of $x$ ($10^{-5}<x<10^{-3}$).  
In the semihard region 
$Q_0^2<Q^2<Q^2_{\rm SH}$,
$\gamma (x,Q^2)$ is large and $R_g$ approaches 
$R_g^{\rm PT}(x,Q^2_{\rm SH})$ fast regardless of what the initial
shadowing is at $Q_0^2$. It is precisely in this region that
$\kappa_g$ is large and (\ref{eq:rptk}) gives a large shadowing correction.
As $Q^2>Q^2_{\rm SH}$, the anomalous dimension $\gamma$ gets small
and in the meantime $R_g^{\rm PT}$ approaches $1$ rather quickly, the
shadowing ratio $R_g$ evolves slowly. Thus, the behavior of 
the $Q^2$-evolution of the gluon shadowing can be understood by the
interplay between the anomalous dimension of the nucleon gluon structure
function and the perturbative shadowing mechanism.

In Fig.\ \ref{fig:rgpt} we show the $Q^2$-behavior of $\gamma (x,Q^2)$ and 
$R_g^{\rm PT}(x,Q^2)$ at $x=10^{-4}$ for $^{208}$Pb. The evolution
of the shadowing ratio
$R_g$ with an initial condition $R_g^0(x)=0.5$ is also plotted to illustrate
the evolution governed by (\ref{eq:rgq}).

To study the sensitivity of the shadowing ratio on the initial condition,
one can solve (\ref{eq:rgq}) explicitly by making a transformation
$R_g=R_g'\exp [-\int \gamma (x,t') dt']$ where $t=\ln Q^2$. Then $R_g'$
can be directly integrated out which yields
\begin{eqnarray}
R_g(x,t) &= & R_g(x,t_0)e^{-\int_{t_0}^t\gamma (x,t')dt'} 
+ e^{-\int_{t_0}^t\gamma (x,t')dt'}\label{eq:sol}\\
& & \times \int_{t_0}^t\gamma (x,t')
e^{\int_{t_0}^{t'}\gamma (x,t'')dt''}R_g^{\rm PT}(x,t')dt'
\; .\nonumber
\end{eqnarray}
In the absence of perturbative shadowing, i.e.\ $R_g^{\rm PT}=1$,
the integral in (\ref{eq:sol}) can be done exactly leading to
a pure GLAP evolution
\begin{eqnarray}
R_g^{\rm GLAP}(x,t) =  1-[1-R_g(x,t_0)]e^{-\int_{t_0}^t\gamma (x,t')dt'}
\; . \label{eq:resol}
\end{eqnarray}
In the DLA, the integration of $\gamma (x,t)$ yields 
\begin{eqnarray}
e^{-\int_{t_0}^t\gamma (x,t')dt'}  = 
\frac{xg_N^{\rm DLA}(x,t_0)}{xg_N^{\rm DLA}(x,t)}
=  \frac{1}{I_0(y)}
\label{eq:damp}
\end{eqnarray}
where $y$ is defined in (\ref{eq:y}). The factor 
$1/I_0(y)$ suppresses the
initial shadowing $R_g(x,t_0)$ quickly for large 
$t>t_{\rm SH}=\ln Q^2_{\rm SH}$. The integration in (\ref{eq:damp})
picks up most important contributions in the region $t_0<t<t_{\rm SH}$.
Therefore, $R_g^{\rm GLAP}\rightarrow 1$ quickly as $t>t_{\rm SH}$.
In the presence of perturbative shadowing, i.e.\ $R_g^{\rm PT}<1$, 
$R_g(x,t)$ at large $t$ is mainly determined by 
the second term in (\ref{eq:sol}) and it depends only on $\gamma $ and
$R_g^{\rm PT}$, especially in the small-$x$ region. 

\begin{figure}
\vskip -0.2in
\centerline{\epsfig{figure=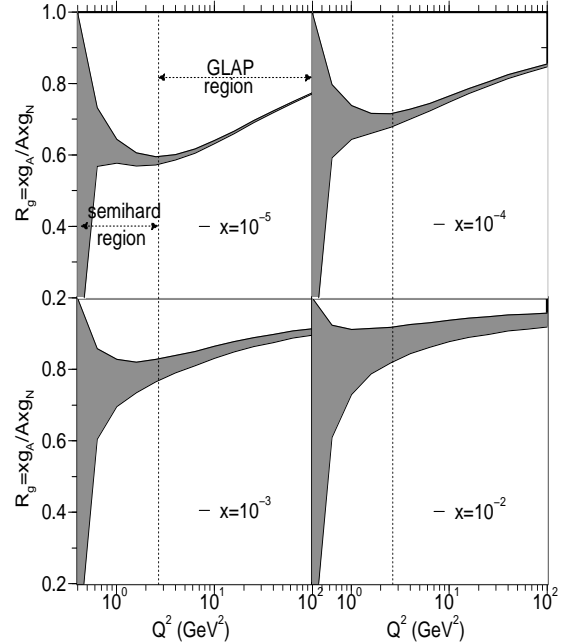,width=3.5in,height=3.5in}}
\caption{The maximal uncertainty in predicting the gluon shadowing
due to initial conditions. The shaded regions are bounded by 
two extreme initial conditions: $R_g^0(x)=1$ and $R_g^0(x)=0$.}
\label{fig:qev}
\end{figure}

Fig.\ \ref{fig:qev} demonstrates the stabilization of the gluon shadowing
for any initial conditions as $Q^2$ increases at various values
of $x$ for $^{208}$Pb. The shaded regions indicate the maximal
uncertainty in predicting the gluon shadowing as they are bounded by
two extreme initial conditions: $R_g^0(x)=1$ and $R_g^0(x)=0$ for all
values of $x$. One can almost
uniquely determine the amount of the gluon shadowing for 
$Q^2>3$ GeV$^2$. In general, the uncertainty gets bigger as $x$
gets larger due to a weaker scaling violation, which is evident
in Fig.\ \ref{fig:qev}. 
Neglecting the first term in (\ref{eq:sol}) and approximating
the perturbative shadowing by 
$R_g^{\rm PT}=R_g^{\rm PT}(Q^2_{\rm SH})$ for 
$Q_0^2<Q^2<Q^2_{\rm SH}$ and $R_g^{\rm PT}=1$ for $Q^2>Q^2_{\rm SH}$,
one obtains from (\ref{eq:sol}) the shadowing ratio for
$Q^2>Q^2_{\rm SH}$
\begin{eqnarray}
R_g(x,t)=1-\left [1-R_g^{\rm PT}(x,t_{\rm SH})\right ]
e^{-\int_{t_{\rm SH}}^t\gamma (x,t')dt'}
\; .\label{eq:psol}
\end{eqnarray}
Comparing (\ref{eq:psol}) with (\ref{eq:sol}), one notices that 
solution  (\ref{eq:psol}) is just solution (\ref{eq:resol}) with
a new initial condition $R_g^0=R_g^{\rm PT}(x,Q^2_{\rm SH})$ and
$Q_0^2$ replaced by $Q^2_{\rm SH}$.
The difference  is that the evolution
factor $\exp [-\int ]$ no longer falls rapidly and $R_g$ evolves
rather slowly. Thus, the $Q^2$-behavior of gluon shadowing at large
$Q^2$ can be roughly described by an initial condition at $Q^2_{\rm SH}$
followed by a GLAP evolution. The important point is that the initial
condition at $Q^2=Q^2_{\rm SH}$ 
is pre-determined by the perturbative shadowing. 
Our main conclusion is that even though 
$R_g^{\rm PT}(x,Q^2)$ approaches $1$ quickly as $Q^2$ increases
beyond $Q^2_{\rm SH}$ (a higher twist effect), 
the shadowing ratio at large $Q^2$ is still governed by the perturbative 
shadowing at the {\em semihard scale}, namely $R_g^{\rm PT}(x,Q^2_{\rm SH})$.

\begin{figure}
\vskip -0.2in
\centerline{\epsfig{figure=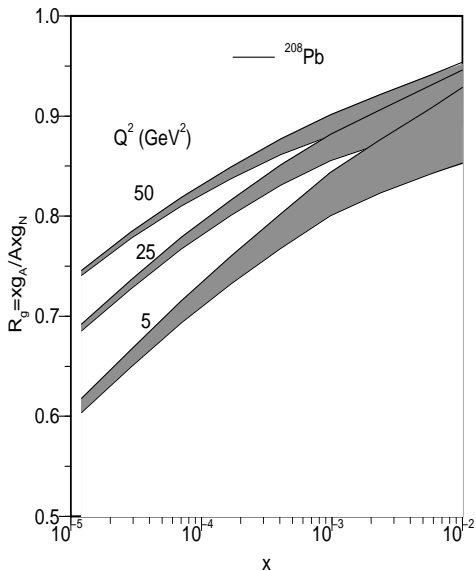,width=3.5in,height=3.5in}}
\vskip -0.2in
\caption{The $x$-dependence of the gluon shadowing
ratio at various $Q^2$ values. The shaded region indicates the 
uncertainty due to the initial conditions.}
\label{fig:xev}
\end{figure}

The $x$-dependence of the gluon shadowing can also be predicted as long
as $Q^2>Q^2_{\rm SH}$ where the influence of the initial condition is 
minimal. Fig.\ \ref{fig:xev} shows the $x$-dependence of the gluon shadowing
ratio at various $Q^2$ values. The shaded region is bounded
by the distributions calculated using two drastically different initial
conditions $R_g^0(x)=0$ and
$R_g^0(x)=1$. It is clear that the shape of the distribution
 is quite robust in the 
small-$x$ region regardless
of what initial conditions one may choose. Due to the perturbative
nature of the shadowing, these distributions do not exhibit a saturation
as $x$ decreases. This is related to the singular behavior of the nucleon
gluon distribution at large $Q^2$ and small $x$. As long as the nucleon
gluon density does not saturates at small $x$, the shadowing ratio does 
not saturate either. 
This is in accordance with the observation by Eskola, Qiu
and Wang  in the context of the gluon recombination model \cite{eqw}.
The rapid $Q^2$-evolution of the gluon shadowing ratio 
in the low $Q^2$ region found by
Eskola \cite{eskola1} can be understood in our approach
as the quick onset of the perturbative shadowing. 
In Eskola's approach some shadowing persists even at $Q^2=100$ GeV$^2$
 which can also be understood
in our picture as the region of fully developed GLAP evolution.
We note that multiple scatterings inside the nucleon are also
possible in the region of very small $x$, where the
eikonalization of  the nucleon cross section is needed.  However,
in this region, the DLA breaks down and a BFKL resummation is necessary.

\section{Gluon Shadowing at Fixed Impact Parameter}
\label{sec:impact}
So far we have confined ourselves to the impact parameter averaged
nuclear structure function. However, in the binary approximation of
hard scattering processes in $pA$ and $AA$ collisions, one has to 
calculate the parton cross section at the nucleon-nucleon level at
each impact parameter. In addition, to enhance the possible signal
of a quark-gluon plasma in heavy-ion collisions, one triggers on 
central events where the impact parameter is small (e.g.\ $b<2$ fm).
It is thus necessary to study the impact parameter dependence of 
the nuclear shadowing effect.

Intuitively, one expects the nuclear shadowing effect
to depend on the local nuclear thickness at a given impact
parameter. In the Glauber approach, the
effective number of scatterings is the product of the cross section and
the nuclear thickness function. Since the shadowing is a non-linear
effect in the effective number of scatterings, the impact parameter dependent
shadowing ratio cannot be factorized into a product of
an average shadowing ratio and the nuclear thickness function \cite{eskola2}.
 \begin{figure}
\vskip 0.3in
\centerline{\epsfig{figure=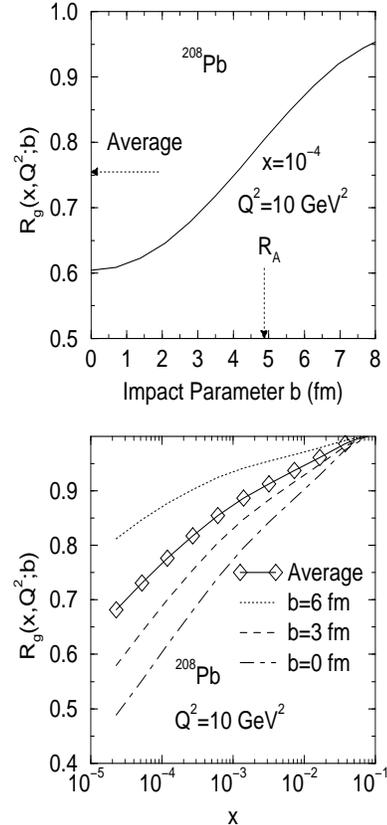,width=4in,height=4in}}
\caption{The impact parameter dependent gluon shadowing
ratio calculated assuming $R_g^0(x,Q_0^2;b)=1$ for any $b$. The upper panel
is the shadowing ratio as function of $b$ at $x=10^{-4}$ and
$Q^2=10$ GeV$^2$. Tle lower panel shows the $x$-dependence of
the shadowing ratio at $Q^2=10$ GeV$^2$ and various
values of $b$.}
\label{fig:imp}
\end{figure}
The impact parameter dependent nuclear structure function
$x\tilde{g}_A(x,Q^2;{\bf b})$ is defined as 
\begin{eqnarray}
 xg_A(x,Q^2)=\int d^2{\bf b} x\tilde{g}_A(x,Q^2;{\bf b})\; .\label{eq:imdef}
\end{eqnarray}
Comparing (\ref{eq:imdef}) with (\ref{eq:glb}), one finds
\begin{eqnarray}
x\tilde{g}_A(x,Q^2;{\bf b})& = & x\tilde{g}_A(x,Q_0^2;{\bf b})+
\frac{4}{\pi^3}\int_x^1\frac{dx'}{x'}\int_{Q_0^2}^{Q^2} dQ'^2  \nonumber\\
& & \times2\left [ 1-e^{-\sigma_{ggN}T_A({\bf b})/2}\right ]\; ,\label{eq:imga}
\end{eqnarray}
where $x\tilde{g}_A(x,Q_0^2;{\bf b})$ is the initial condition at fixed
impact parameter ${\bf b}$. As shown in the previous section,
the prediction for gluon shadowing ratio at $Q^2\ge 3$GeV$^2$ does not
depend on the initial condition. We take the initial shadowing
ratio  to be $1$ for all ${\bf b}$'s, i.e.\ no initial shadowing. 
The impact parameter dependent
shadowing ratio is thus defined 
\begin{eqnarray}
R_g(x,Q^2;{\bf b})=\frac{x\tilde{g}_A(x,Q^2;{\bf b})}{T_A({\bf b})xg_N(x,Q^2)}
\; . \label{eq:imrg}
\end{eqnarray}

In Fig.\ \ref{fig:imp} we show the gluon shadowing at 
various impact parameters.
The upper panel is the shadowing ratio as function of $b$ at 
$x=10^{-4}$ and $Q^2=10$ GeV$^2$. The parton density is more
suppressed at a small impact parameter. The $x$-dependence of the
impact parameter dependent shadowing ratio is also shown in the lower
panel of Fig.\ \ref{fig:imp}. 

\section{Conclusions}
\label{sec:conc}
We have studied the nuclear shadowing mechanism in the context of
perturbative QCD and the Glauber multiple scattering model. 
The nuclear shadowing phenomenon is a consequence of the
coherent parton multiple scatterings.
While the quark density shadowing arises from an interplay between 
the ``soft''
and the semihard QCD processes, the gluon shadowing is 
largely driven by a perturbative shadowing mechanism due to the strong
scaling violation in the small-$x$ region. 
The gluon shadowing is thus a robust phenomenon at large $Q^2$ and can
be unambiguously predicted by perturbative QCD.

The scale dependence of the nucleon structure function 
has been used to put a stringent constraint on
the gluon density inside the nucleon in the double logarithm approximation.
It is shown that 
the strong scaling violation of the nucleon structure function in the
semihard momentum transfer region $0.8\le Q^2 \le 3$ GeV$^2$
at small $x$ can be reliably described
by perturbative QCD and 
is  a central key to the
understanding of the scale dependence of the nuclear shadowing effect.

We have also studied
the impact parameter dependence of gluon shadowing and shown that 
it is a non-linear effect in the nuclear thickness function. It is
important to correctly incorporate the impact parameter dependence
of the nuclear structure function when one calculates the QCD processes
such as minijet production and J/$\psi$ production 
in the central nucleus collisions. 

\acknowledgments

This work was supported 
through the U.S. Department of Energy under Contracts Nos.\ 
DE-FG03-93ER40792 and DE-FG03-85ER40213.

\appendix
\begin{center}
{\small \bf APPENDIX: PARAMETRIZATION OF GLUON \\ SHADOWING AND ITS
IMPACT \\ PARAMETER DEPENDENCE}
\end{center}
Motivated by our studies of the gluon shadowing at $Q^2>Q_{\rm SH}^2$,
we parametrize the gluon shadowing ratio in the following way:
\begin{eqnarray}
R_g(x,Q^2) & = & 1-[1-R_g^{\rm PT}(x,Q_{\rm SH}^2)]\nonumber \\
& & \times\left [ 
1-a_0\ln {Q^2}/{Q_{\rm SH}^2}\right ]\; ,
\end{eqnarray}
where $Q_{\rm SH}^2=3$ GeV$^2$, $a_0=0.3$ and
\begin{eqnarray}
R_g^{\rm PT}(x,Q_{\rm SH}^2)=1-a_1A^\alpha x^{-\lambda}(1-x)^\beta \; ,
\end{eqnarray}
where $A$ is the atomic number of the nucleus and
\begin{eqnarray}
a_1=0.0249\; ; \quad \alpha =0.18\; ; 
\quad \lambda =0.162\; ; \quad \beta =18\; .
\end{eqnarray}

The impact parameter dependent ratio $R_g(x,Q^2;{\bf b})$
is parametrized via a $b$-averaged ratio $R_g(x,Q^2)$
\begin{eqnarray}
R_g(x,Q^2;{\bf b}) & = & 1-c_0(1+c_1A^{-\alpha '}b^2) \\
& & \times\exp [-A^{-\alpha '}b^2/c_2]
(1-R_g(x,Q^2))\nonumber\; ,
\end{eqnarray}
where $b$ is the impact parameter in fm and
\begin{eqnarray}
c_0=1.7\; ; \; c_1 =0.043\; ; 
\; \alpha ' =0.11\; ; \; c_2 =11.45\; .
\end{eqnarray}
Note that these parametrizations are valid for $Q^2>Q_{\rm SH}^2$
and $x<10^{-2}$.


\begin{references}
\bibitem{cern}NMC Collaboration, P.\ Amaudruz {\it et al.}, Z.\ Phys.
C {\bf 51}, 387 (1991); Z.\ Phys.C {\bf 53}, 73 (1992); Phys.\ Lett.\ 
B {\bf 295}, 195 (1992); NMC Collaboration, M. Arneodo {\it et al.},
Nucl.\ Phys.\ B {\bf 441}, 12 (1995).
\bibitem{fermilab}E665 Collaboration, M.R.\ Adams {\it et al.},
Phys.\ Rev.\ Lett.\ {\bf 68}, 3266 (1992); Phys.\ Lett.\ B {\bf 287},
375 (1992); Z. Phys. C {\bf 67}, 403 (1995).
\bibitem{glushdw}D.\ Alde {\it et al.}, Phys.\ Rev.\ Lett. {\bf 66}, 
133 (1991).
\bibitem{fs}S.\ Gavin and J.\ Milana, Phys.\ Rev.\ Lett. {\bf 68},
1834; C.J.\ Benesh, J.\ Qiu and J.P.\ Vary, Phys.\ Rev.\ C {\bf 50},
1015 (1994).
\bibitem{rev}B. Badelek and J. Kwieci\'{n}ski, Rev.\ Mod. Phys.
{\bf 64}, 927 (1992); M. Arneodo, Phys. Rep. {\bf 240}, 301 (1994).
\bibitem{glr}L.V. Gribov, E.M. Levin and M.G. Ryskin, Phys. Rep.
{\bf 100}, 1 (1983); E. Laenen and E. Levin, Nucl. Phys. B {\bf 451}, 207
(1995).
\bibitem{mq}A.H. Mueller and J. Qiu, Nucl. Phys. B {\bf 268}, 427 (1986);
J. Qiu, Nucl. Phys. B {\bf 291}, 746 (1987).
\bibitem{eqw}K.J. Eskola, J. Qiu and X.-N. Wang, Phys. Rev. Lett.
{\bf 72}, 36 (1994).
\bibitem{frankfurt}L. Frankfurt and M. Strikman, Nucl. Phys. B {\bf 316},
340 (1989); L. Frankfurt, M. Strikman and S. Liuti, Phys. Rev. Lett.
{\bf 65}, 1725 (1990).
\bibitem{lu}S.J. Brodsky and H.J. Lu, Phys. Rev. Lett. {\bf 64}, 1342 (1990).
\bibitem{zakharov}N.N. Nikolaev, Z. Phys. C {\bf 32}, 537 (1986);
N.N. Nikolaev and B.G. Zakharov, Phys. Lett. B {\bf 260},
414 (1991); Z. Phys. C {\bf 49},
607 (1991).
\bibitem{mueller}A.H. Mueller, Nucl. Phys. B {\bf 335}, 115 (1990).
\bibitem{agl}A.L. Ayala F, M.B. Gay Ducati and E.M. Levin, 
e-Print Archive: hep-ph/9604383.
\bibitem{glap}V.N. Gribov and L.N. Lipatov, Sov. J. Nucl. Phys. {\bf 15},
438 (1972); G. Altarelli and G. Parisi, Nucl. Phys. B {\bf 126}, 298 (1977);
Yu.L. Dokshitzer, Sov. Phys. JETP {\bf 46}, 641 (1977).
\bibitem{vmd}see e.g., G. Piller, W. Ratzka and W. Weise, Z. Phys. 
A {\bf 352}, 427 (1995); 
T.H. Bauer, R.D. Spital and D.R. Yennie and F.M.\ Pipkin,
Rev. Mod. Phys. {\bf 50}, 261 (1978).
\bibitem{eskola1}K.J. Eskola, Nucl. Phys. B {\bf 400}, 240 (1993).
\bibitem{mrs}A.D. Martin, R.G. Roberts and W.J. Stirling, Phys. Lett. 
B {\bf 387}, 419 (1996); H.L. Lai, J. Huston, S. Kuhlmann, F. Olness, 
J. Owens, D. Soper, W.K. Tung and H. Weerts, Phys. Rev. D {\bf 55},
1280 (1997); M. Glueck, E. Reya and A. Vogt, Z. Phys. C {\bf 67}, 433 (1995).
\bibitem{glauber}R.J.\ Glauber, in {\it Lectures in theoretical physics},
ed. W.E. Brittin {\it et al.} (Interscience Publishers, New York, 1959);
R.J.\ Glauber and G.\ Matthiae, Nucl. Phys. B {\bf 21}, 135 (1970).
\bibitem{bfkl}E.A. Kuraev, L.N. Lipatov and V.S. Fadin, Sov. Phys. JETP
{\bf 45}, 199 (1977); Ya.Ya. Balitskii and L.V. Lipatov, Sov. J. Nucl. Phys.
{\bf 28}, 822 (1978); L.N. Lipatov, Sov. Phys. JETP {\bf 63}, 904 (1986).
\bibitem{mv}L. McLerran and R.\ Venugopalan, Phys. Rev. D {\bf 49}, 2233
(1994); A. Kovner, L. McLerran and H. Weigert, Phys. Rev. D {\bf 52}, 
6231 (1995); D {\bf 52}, 3809 (1995).
\bibitem{plc}A.J. Brodsky and A.H. Mueller, Phys. Lett. B {\bf 206}, 685
(1988); G. Farrar, L. Frankfurt, M. Strikman and H. Liu, Phys. Rev. Lett.
{\bf 64}, 1996 (1990).
\bibitem{dla}For recent studies, see 
R.K. Ellis, Z. Kunszt and E.M. Levin, Nucl. Phys. B {\bf 420}, 517 (1994);
 R.D. Ball and  S. Forte, Phys. Lett. B {\bf 335}, 77 (1994). 
\bibitem{strikman} B. Blattel, G. Baym, L. Frankfurt and M. Strikman,
Phys. Rev. Lett. {\bf 71}, 896 (1993); L. Frankfurt, G. Miller and 
M. Strikman, Phys. Lett. B {\bf 304}, 1 (1993).
\bibitem{h1}H1 Collaboration, T. Ahmed {\it et al.}, Nucl. Phys. B
{\bf 470}, 3 (1996); H1 Collaboration ,C. Adloff {\it et al.}, 
DESY-97-042, e-Print Archive: hep-ex/9703012.
\bibitem{landshoff}A. Donnachie and P.V. Landshoff, Phys. Lett.
B {\bf 191}, 309 (1987); Z. Phys. C {\bf 61}, 139 (1994).
\bibitem{boris}B. Kopeliovich and B. Povh, Phys. Lett. B {\bf 367}, 
329 (1996). 
\bibitem{eskola2}K.J. Eskola, Z. Phys. C {\bf 51}, 633 (1991).
\end{references}
\end{document}